# Berry curvature induced anomalous Hall conductivity in magnetic topological oxide double perovskite Sr$_2$FeMoO$_6$

Tirthankar Chakraborty[*,1,2] Kartik Samanta,[1] Satya N. Guin,[1,3] Jonathan Noky,[1] Iñigo Robredo,[1,4] Suchitra Prasad,[1] Juergen Kuebler,[5] Chandra Shekhar,[1] Maia G. Vergniory[*,1,4] and Claudia Felser[*1]

[1)]Max Planck Institute for Chemical Physics of Solids, 01187 Dresden, Germany
[2)]School of Physics and Material Science, Thapar Institute of Engineering and Technology, Punjab-147001, India[a]
[3)]Department of Chemistry, Birla Institute of Technology and Science Pilani, Hyderabad Campus, Hyderabad 500078, India
[4)]Donostia International Physics Center, 20018 Donostia - San Sebastián, Spain
[5)]Technische Universitaet Darmstadt, 64289 Darmstadt, Germany

Oxide materials exhibit several novel structural, magnetic, and electronic properties. Their stability under ambient conditions, easy synthesis, and high transition temperatures provide such systems with an ideal ground for realizing topological properties and real-life technological applications. However, experimental evidence of topological states in oxide materials is rare. In this study, we have synthesized single crystals of oxide double perovskite Sr$_2$FeMoO$_6$ and revealed its topological nature by investigating its structural, magnetic, and electronic properties. We observed that the system crystallized in the cubic space group Fm$\bar{3}$m, which is a half-metallic ferromagnet. Transport measurements show an anomalous Hall effect, and it is evident that the Hall contribution originates from the Berry curvature. Assuming a shift of the Fermi energy towards the conduction band, the contribution of the anomalous Hall effect is enhanced owing to the presence of a gaped nodal line. This study can be used to explore and realize the topological properties of bulk oxide systems.

## I. INTRODUCTION

Topological materials are a class of quantum matter that possesses a nontrivial band topology[1–8]. The anomalous Hall effect (AHE) is one of the unique transport phenomena in which nontrivial topological state is discernible. The extrinsic mechanism of AHE originates from skew-scattering or side jump of electrons due to spin-orbit coupling[9,10]. The intrinsic contribution can be obtained from the sum of the Berry curvature of the Bloch wave function over occupied electronic states[11-13]. In a ferromagnetic topological system, the magnetic ordering breaks the time-reversal symmetry, which lifts the degeneracy and can open up a gap in the nodal lines. If such gaped nodal lines are near the Fermi energy, they may lead to a strong Berry curvature that enhances the anomalous Hall conductivity (AHC)[14,15]. Extensive studies of such phenomena and various other topological features have been carried out mostly on Heusler, chalcogen, pnictogen, and pyrite compounds[16–26]. In addition, there is a distinct class of materials, namely oxides, which remain interesting for versatile properties to date. One of the most interesting families is perovskite (or double perovskite) with the chemical formula ABO$_3$ (or A$_2$BB'O$_6$), where, A is generally an alkali metal or rare-earth cation and, B and B' are transition metal elements. Almost all choices of ferromagnetic, ferrimagnetic, or antiferromagnetic, dielectric, ferroelectric, metallic, semiconducting, and insulating materials can be found in this family, and their properties can be easily tuned by various combinations of A, B, and B' cations[27–31]. Moreover, they grow easily, are extremely stable and robust in ambient air, and possess a high magnetic ordering temperature[32–36]. Whereas most of the conventional topological materials studied thus far suffer from low transition temperatures, resulting in challenges for their application in potential technologies, a possible solution might be hidden in the area of oxides with high transition temperatures. In addition, some candidates are half-metallic ferromagnets in which conduction electrons are spin-polarized and therefore distinct from conventional ferromagnetic materials. These properties make oxide perovskites and double perovskites excellent candidates for utilizing topological properties in real-life applications such as spintronics, low-dissipation devices, and quantum computing[37–41]. Despite extensive exploration of their electronic and magnetic properties, it is surprising that experimental realization of the topological properties of these materials in the bulk phase is scarce, except for a few theoretical suggestions[42–47]. Sr$_2$FeMoO$_6$ (SFMO) is a well-known ferromagnetic double-perovskite system[48,49] and has been predicted to be half metallic[48,50]. Investigation of the magnetotransport properties revealed anomalous Hall conductivity in SFMO; however, its origin and whether it is intrinsic in nature have not been explored yet[49]. In this study, we successfully grew single crystals of SFMO and investigated their topological nature through experimental and theoretical studies. The structure of the system is described with the cubic space group Fm$\bar{3}$m, which is a half-metallic ferromagnet. The system shows an anomalous Hall effect, and we reveal that the AHE has an intrinsic contribution owing to the Berry curvature, which can be enhanced by doping owing to the presence of a gaped nodal line above the Fermi energy.

[a] Electronic mail: tirtha255@gmail.com

## II. METHODOLOGY

Single-crystal growth of SFMO was facilitated using a four-mirror laser floating-zone furnace. The growth was carried out at a rate of 10 mm/h in an Ar atmosphere of pressure 0.2 MPa. A polycrystalline ceramic rod prepared using a pure phase polycrystalline powder sample was used as the feed rod. The detailed process of the preparation of the polycrystalline sample in the pure phase will be discussed elsewhere. During the growth, the feed and seed rods were counter-rotated at 30 rpm to obtain temperature and chemical homogeneity. The grown crystals were oriented using the OrientExpress software program on the Laue back-reflected patterns and cut using a wire saw. Magnetic measurements were performed on the single crystals using a SQUID MPMS-3 magnetometer (Quantum Design). The electrical transport properties were measured using a physical property measurement system (PPMS9) (Quantum Design). For transport measurements, the crystals were cut into standard rectangular shapes, and a six-probe technique was used to measure the longitudinal and hall resistivity simultaneously. The recorded data were symmetrized (anti-symmetrized) to eliminate electrode misalignment.

Electronic structure calculations were performed using the density functional theory (DFT)-based planewave projected augmented wave (PAW) method, as implemented in the Vienna ab initio Simulation Package (VASP)[51–53]. For the electronic structure calculations, we used the experimentally determined cubic symmetry (Fm$\bar{3}$m) of SFMO, with a lattice parameter of 7.89 Å. We used a 12 × 12 × 12 k-point mesh and a plane wave cutoff of 520 eV for the self-consistent calculations, which provided good convergence of the total energy. We used the Perdew-Burke-Ernzerhof (PBE)[54] exchange correlation functional within the generalized gradient approximation (GGA). The electron-electron correlation effects beyond GGA at the magnetic sites were taken into account by the onsite Coulomb interaction strength $U$, and the intra-atomic exchange interaction strength $J_H$ [55], chosen to be 4.0, 2.1, and 0.8, 0.45 eV for Fe−3$d$ and Mo−5$d$, respectively which were recently estimated using the constrained random phase approximation (cRPA)[56,57].

Employing the Wannier interpolation technique[58], we assess the intrinsic Berry curvature contribution to the anomalous Hall conductivity. To compute the Berry curvature, we first constructed a maximally localized Wannier function (MLWFs) Hamiltonian projected from the GGA+$U$+SOC Bloch wave functions using the Wannier90 package[59–61]. Using a numerical tight-binding model Hamiltonian, we calculated the intrinsic AHC using the linear response Kubo formula approach, as follows[62]:

$$\Omega_n^z(\mathbf{k}) = -\hbar^2 \sum_{n \neq m} \frac{2\mathrm{Im}\langle u_{m\mathbf{k}}|\hat{v}_x|u_{n\mathbf{k}}\rangle \langle u_{n\mathbf{k}}|\hat{v}_y|u_{m\mathbf{k}}\rangle}{(\epsilon_{m\mathbf{k}} - \epsilon_{n\mathbf{k}})^2} \quad (1)$$

where $\Omega_n(\mathbf{k})$ is the Berry curvature of band $n$, $\hbar \hat{v}_i = \partial \hat{H}(\mathbf{k})/\partial k_i$ is the $i$'th velocity operator, and $u_{n\mathbf{k}}$ and $\epsilon_{n\mathbf{k}}$ are the eigenstates and eigenvalues of Hamiltonian $\hat{H}(\mathbf{k})$, respectively.

Subsequently, we calculated the AHC, given by:

$$\sigma_H^A = -e^2/\hbar \sum_n \int_{BZ} \frac{d\mathbf{k}}{(2\pi)^3} \Omega_n^z(\mathbf{k}), \quad (2)$$

We used a k-point mesh of size 300 × 300 × 300 to calculate the AHC using Eq. 2.

## III. RESULTS AND DISCUSSION

### A. Structure and Magnetic properties

The room-temperature powder X-ray diffraction pattern obtained for SFMO is shown in Fig. 1. The experimental data were refined by the Rietveld method using the FullProf software. The structure can be refined using cubic space group Fm$\bar{3}$m (225) with a lattice parameter of $a$ = 7.890 Å. This is consistent with the optimized structure obtained using ab initio calculations in the same space group[63].

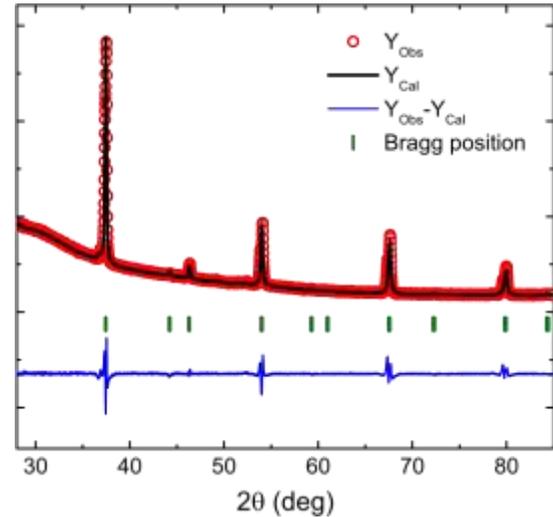

FIG. 1. (Color online) Powder X-ray diffraction pattern (Y$_{obs}$) obtained at 300 K is shown along with the calculated pattern (Y$_{cal}$) obtained by Rietveld analysis using space group Fm$\bar{3}$m. The difference pattern and the Bragg positions are also shown. No sign of any secondary phase is observed.

Magnetic isotherms at various temperatures with magnetic fields along the B∥a and B∥c axes are shown in Fig. 2a and 2b, respectively. In both crystallographic directions, the hysteresis loops are almost isotropic, which is expected for a cubic structure. The saturation magnetization at 2 K was 2.87 $\mu_B/f.u.$, which is consistent with a previously reported value[49]. To understand the saturation magnetic moment from the DFT calculations, double perovskites[64,65]. The Hall resistivity $\rho_{yx}$ first increases up to magnetic field $B$ ($B = \mu_0 H$) by approximately 1 T, which is attributed to the anomalous Hall effect. However, it decreases with a further increase in the field above 1 T. Corresponding Hall conductivity ($\sigma_{xy}$) is given by $\sigma_{xy} = \frac{\rho_{yx}}{\rho_{yx}^2 + \rho_{xx}^2}$, which shows a similar dependence on the field as $\rho_{yx}$ and, as shown in Fig. 3d.

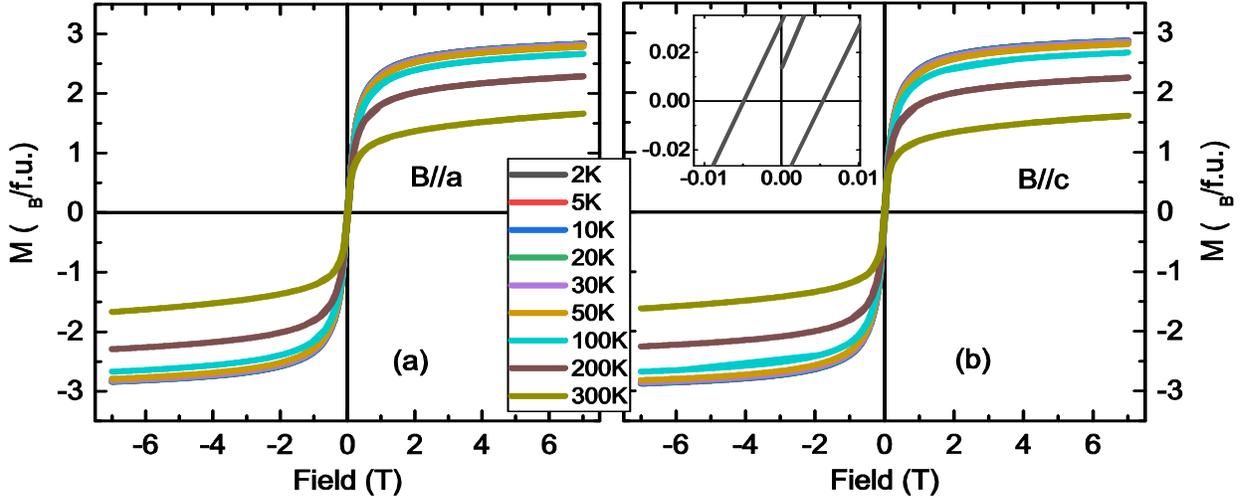

FIG. 2. (Color online) Isothermal magnetization at various temperatures with magnetic field along a and c axis are shown in (a) and (b) respectively. Inset of (b) shows the magnified hysteresis curve at 2 K.

we calculated the localized magnetic moment at the Fe–3$d$ and Mo–5$d$ site and found it to be 4.06 and -0.54 $\mu_B$, respectively, consistent with the half-filled Fe-3d$^5$ [t$_{2g}$(3↑), e$_g$(2↑)] and Mo-5d$^1$ [t$_{2g}$(1↓)] states. A net magnetization of 4 $\mu_B/f.u.$ was obtained from DFT calculations, as expected in the atomic limit considering Fe in the 3d$^5$ (S=5/2) and Mo in the 5d$^1$ (S=1/2) states. The discrepancy between the experimentally measured magnetization value of 2.87 $\mu_B/f.u.$, and a computed value of 4.0 $\mu_B/f.u.$ can be attributed to the presence of anti-site disorder between Fe and Mo sites, whereas in first-principles calculations, we consider the perfect ordering of Fe and Mo sites.

**B. Transport and magnetotransport properties**

The electric transport and magnetic field-dependent Hall resistivity $\rho_{yx}$, longitudinal resistivity $\rho_{xx}$, and Hall conductivity $\sigma_{xy}$ at several temperatures are shown in Fig. 3. Similar to the magnetic properties, the transport properties were also observed to be isotropic with a magnetic field along different crystallographic directions. The temperature dependence of the resistivity shows a typical metallic behavior. With increasing magnetic field, $\rho_{xx}$ decreases, and the observed magnetoresistance is −14.7% at 2 K under a maximum field of 9 T, which is moderate and comparable to that of many other similar

The Observed $\rho_{yx}$ consists of two contributions from a field-dependent ordinary term and an anomalous term:

$$\rho_{yx} = \rho_{yx}^O(B) + \rho_{yx}^A(M) \quad (3)$$

where $B$ is the magnetic induction, $\rho_{yx}^O$ and $\rho_{yx}^A$ are the ordinary and anomalous Hall resistivities, respectively. The ordinary Hall coefficient ($R_H$) corresponding to $\rho_{yx}^O$ can be obtained from $d\rho_{yx}/dH$ in the high-field region where $M$ has already reached saturation. A negative slope, as shown in Fig. 3c within the region $\mu_0 H > 1$ T, indicates that the dominant charge carrier is electron-like. The corresponding carrier concentration ($n$) can be calculated from the relationship $R_H = 1/ne$, where $e$ is the electron charge. At 2 K, $n$ is of the order of $1.08 \times 10^{22}$ /cm$^3$, which increases exponentially as the temperature increases, as shown in Fig. 4. The anomalous Hall resistivity $\rho_{yx}^A$ involves both intrinsic and extrinsic mechanisms. It can be written as,

$$\rho_{yx}^A = (\alpha \rho_{xx} + \beta \rho_{xx}^2) M(T) \quad (4)$$

where the first term on the right-hand side is due to skew scattering, which is an extrinsic mechanism and linear to $\rho_{xx}$, whereas the second term represents the intrinsic Berry phase and side jump contributions to the AHE, and both vary quadratically with $\rho_{xx}$. $\alpha$ and $\beta$ are two parameters that can be obtained by plotting $\rho_{yx}^A/M\rho_{xx}$ vs.

$\rho_{xx}$. Once $\alpha$ and $\beta$ are extracted, the contributions of skew scattering and intrinsic plus-side jump to $\rho_{yx}^A$ can be calculated. Corresponding AHC due to extrinsic skew scattering ($\sigma_{xy}^{sk}$) and combined effect of intrinsic Berry phase contribution and side jump ($\sigma_{xy}^{int}$) were obtained within the temperature range of 2–50 K, and their variations with temperature are shown in Fig. 5. The highest values of $\sigma_{xy}^{sk}$ and $\sigma_{xy}^{int}$ at 2 K were 59.34 and 46.75 S cm$^{-1}$, respectively and both correspond to anomalous Hall angle of 1.5%.

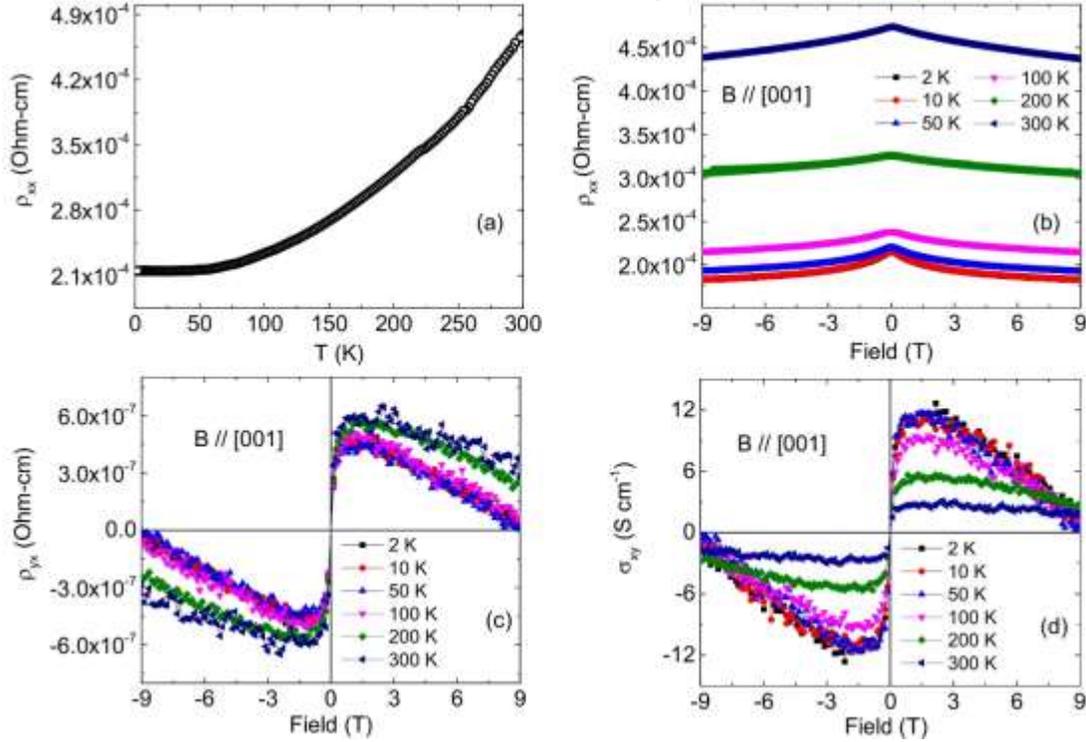

FIG. 3. (Color online) (a) Temperature as a function of resistivity. Variation of longitudinal resistivity $\rho_{xx}$, Hall resistivity $\rho_{yx}$ and Hall conductivity $\sigma_{xy}$ with magnetic field are shown in (b), (c) and (d) respectively at several temperatures.

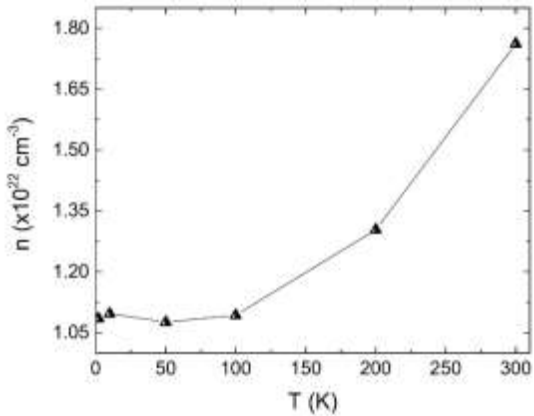

FIG. 4. (Color online) Temperature dependence of carrier concentration obtained from the ordinary Hall coefficient $R_H$ at different temperatures.

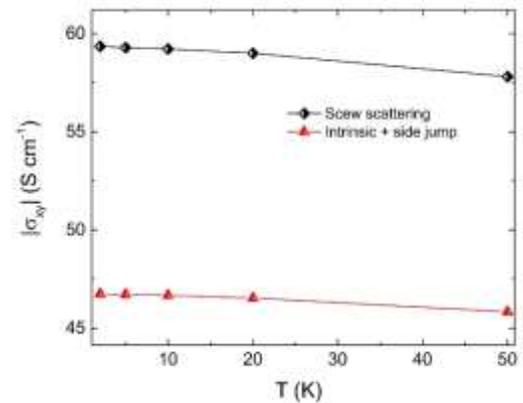

FIG. 5. (Color online) Variation of the anomalous Hall conductivity as a function of the temperature till 50 K. The extrinsic part of the anomalous Hall conductivity ($\sigma_{xy}^{sk}$) originated from skew scattering and intrinsic part of anomalous Hall conductivity ($\sigma_{xy}^{int}$) resulting from Berry phase and side jump are separately shown here.

**C. Computed anomalous Hall conductivity and Nodal line structure**

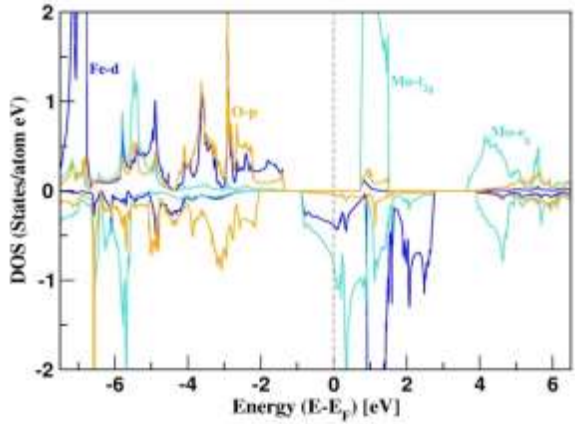

FIG. 6. Spin-polarized density of states, projected onto Fe −$d$ (blue line), Mo−$d$, (cyan line), O−$p$ (orange line) states.

The DFT-calculated spin-polarized density of states (DOS) without the SOC is shown in Fig. 6. Both Fe and Mo are in the regular octahedral environment of oxygen ions and within the crystal field of the regular octahedron, the d levels split into t$_{2g}$ and e$_g$ states. As shown in Fig. 6, the states close to the Fermi level (E$_F$) were dominated by partially filled Mo−$t_{2g}$ states in the down-spin channel. The majority spin channel is filled with Fe−$d$ states, which have finite hybridization with O−$p$ states. The partially filled t$_{2g}$ states of Mo cross the Fermi level in the minority spin channel, which makes the minority spin channel conduct, while there is a clear gap in the majority spin channel, leading to a half-metallic ground state. The ground state remains a half-metallic ferromagnet in the presence of spin-orbit coupling (SOC).

The electronic band structure considering the spin moment aligned along the z-direction with SOC included is shown in Fig. 7a. To compute the intrinsic contribution of AHC, we constructed a maximally localized Wannier function Hamiltonian projected from the Fe−$d$, Mo−$d$, Sr−$d$, and O−$p$ Bloch states, as shown in the right panel of Fig. 7a.

The intrinsic contribution of the AHC at the Fermi energy was calculated to be approximately $\sigma_H^A$=22 S/cm, which is in qualitative agreement with the experimentally measured value of $\sigma_{xy}^{int}$. Small oxygen nonstoichiometry is commonly observed in perovskite oxides[66-68]. Oxygen vacancies appear during the cooling process of the compounds after they are annealed at a high temperature[69,70]. This can effectively uplift the Fermi level, and a small increase in the experimental AHC compared with the theoretical one can be anticipated.

From the computed value of the AHC as a function of the Fermi energy, we observed two large peaks of the AHC away from the Fermi energy (cf. right panel of Fig. 7a). The first peak of the AHC was observed at $E$ = +259 meV and the second peak was observed at $E$ = +365 meV above the Fermi level. As a consequence, a shift of the Fermi energy by 259 meV (which corresponds to 0.54 additional electrons per formula unit) would have a large effect on the observed value of the AHC, which can reach up to approximately 313 S/cm (green line in Fig. 7a). In the following section, we investigate the peaks in the AHC in detail. We found a crossing along the $K$−Γ high-symmetry line at the same energy as the AHC peak, which is surrounded by an orange circle in Fig. 7a. In Fig. 7(b) and (c), we show the 3D band structure in the presence of SOC, considering only the two bands that form the nodal line in planes $k_z$=0 and $k_x$=0. It can be observed that while the first one has no gap, the latter plane hosts a gaped line due to the presence of both SOC and magnetism. Without SOC, the internal spin space and lattice space are decoupled from each other, and spins can be separated into spin-up and spin-down sectors. Therefore, without SOC, the cubic SFMO possesses three mirror planes at $k_z$=0, $k_y$=0, *and* $k_x$=0, and in the magnetic ground state, we find a symmetry-protected nodal line in each of the three planes along the $K$−Γ direction. The situation is different when we consider SOC, which couples the up and down spin channels with each other, and consequently, the spin-rotational symmetry is broken. In particular, when the spin moment is aligned along the z-direction (001), the mirror symmetries M$_x$ and M$_y$ are broken, leading to a gaped nodal line along the $K$ −Γ high-symmetry line in the $k_x$ = 0 plane (Fig. 7c). Instead, for the moment aligned along the z-direction, the mirror symmetry M$_z$ is still preserved, which leads to a gapless nodal line in the $k_z$ = 0 plane (cf. Fig. 7b)[71-73]. These gaped nodal lines are the origin of large Berry curvature in the Brillouin zone and consequently enhance AHC.

To further investigate the contribution of these gaped nodal lines to AHC, we calculated the Berry curvature contribution in these planes. Due to the presence of mirror symmetry at the $k_z$=0 plane, we observed a closed nodal line, which led to an almost zero Berry curvature contribution in the Brillouin zone. However, in the $k_x$ = 0 plane, where mirror symmetry is broken due to magnetization along the z-direction (001), the nodal lines are gaped out, as shown in Fig. 8a, in which case, we can observe a finite contribution to the Berry curvature, as shown in Fig. 8b. To quantify the contributions of the gaped nodal lines to the total AHC, we restricted the integration from Eq. 2 to slices around the $k_x$ = 0 and $k_y$ = 0 planes because they host the gaped lines. When setting the thickness of the slices to 20% of the full BZ, we found 67% of the total AHC in the restricted volume; however, an increase of 30% in thickness yielded a fraction of 87%. From this, we conclude that the gaped nodal lines are the main source of AHC.

## IV. CONCLUSION

In summary, we grew single crystals of SFMO double perovskite and investigated its magnetic and transport properties in detail through experimental and theoretical studies. The structure can be described by the cubic space group Fm$\bar{3}$m, and the ground state was found to be ferromagnetic (half) metal. The transport and magnetic properties are essentially isotropic, as expected for a cubic system. We found that the Berry curvature contributions could reproduce the measured intrinsic contribution of the AHC. In addition, we observed a large AHC contribution of 313 S/cm at 259 meV above the Fermi level. This large AHC value can be attributed to the presence of a gaped nodal line structure, which enhances the Berry curvature. Our work should stimulate further studies to investigate and realize the novel topological properties of similar oxide materials.

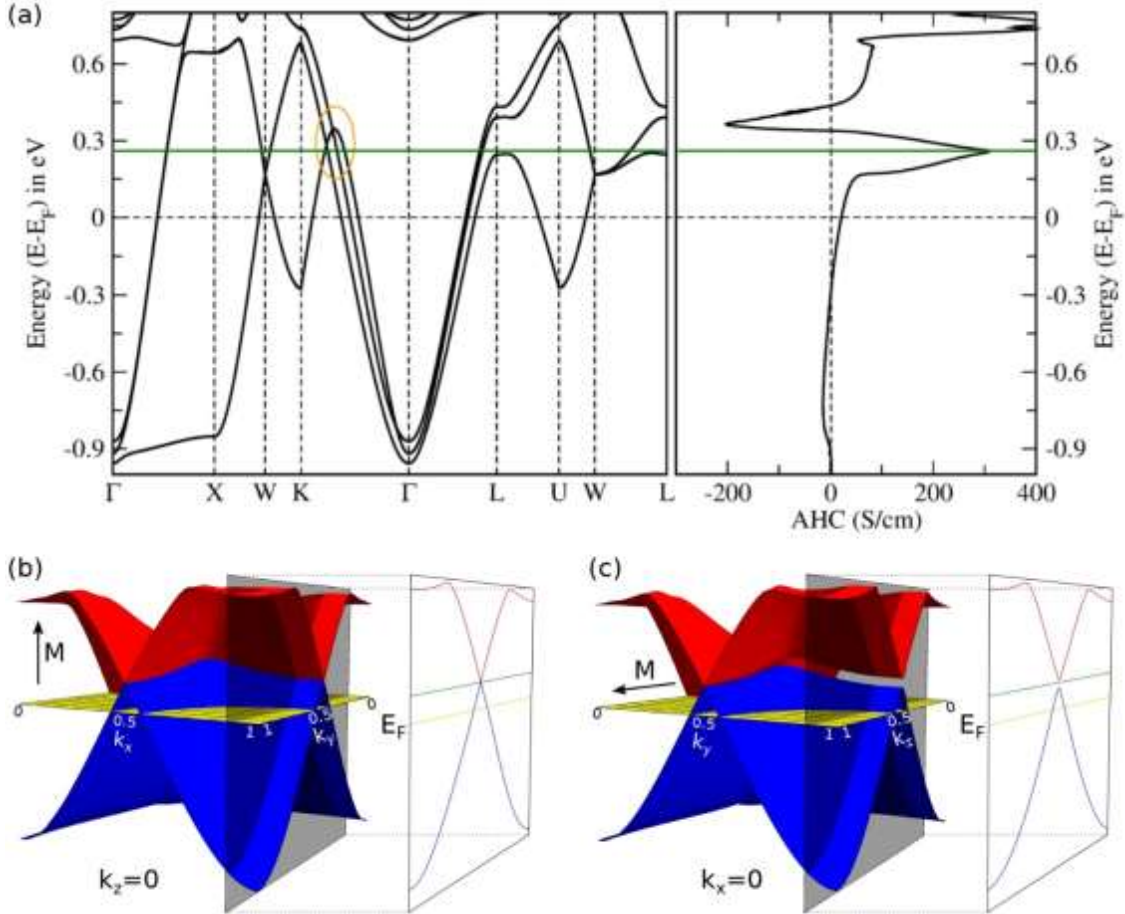

FIG. 7. (a) (Left panel) Band structure of $Sr_2FeMoO_6$ magnetic ground state considering the magnetization M ∥(001). (Right panel) Computed intrinsic AHC contribution. The green line marks the peak in the AHC at $E$ = 259 meV. (b) 3D band structure taking into account the two bands that form the nodal line in the $k_x$-$k_y$ and (c) $k_y$-$k_z$ planes. M is the direction of the fully polarized magnetic moments with SOC. The insets show a cut in the respective planes, showing the closed (b) and gaped (c) nodal lines, depending on the alignment of the magnetic moments with the mirror planes (M ∥ (001)).

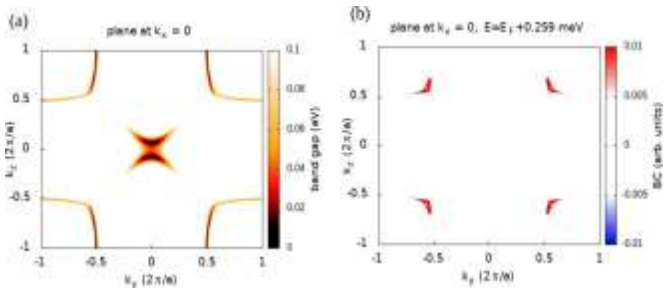

FIG. 8. (a) Band gap of the two bands forming the nodal line in the $k_x$ = 0 plane with magnetization, M∥ (001). (b) Berry curvature in the $k_x$ =0 plane with M∥ (001).

## V. ACKNOWLEDGEMENTS

This work is financially supported by the European Research Council (ERC Advanced Grant No. 742068 'TOPMAT'). We also acknowledge funding from the DFG through SFB 1143 (project ID. 247310070) and the Würzburg-Dresden Cluster of Excellence on Complexity and the Topology in Quantum Matter ct.qmat (EXC2147, project ID. 390858490). M.G.V. thanks support to Programa Red Guipuzcoana de Ciencia Tecnología e Innovación 2021 No. 2021-CIEN-000070-01 Gipuzkoa Next. M.G.V. and I.R. acknowledge Spanish Ministerio de Ciencia e Innovacion (grant PID2019-109905GBC21). M.G.V and C.F. are supported by the Deutsche Forschungsgemeinschaft (DFG, German Research Foundation) FOR 5249 (QUAST).

## VI. REFERENCES

[1] Y. Deng, Y. Yu, M. Z. Shi, Z. Guo, Z. Xu, J. Wang, X. H. Chen, and Y. Zhang, Science **367**, 895 (2020).

[2] C.-Z. Chang, W. Zhao, D. Y. Kim, H. Zhang, B. A. Assaf, D. Heiman, S.-C. Zhang, C. Liu, M. H. Chan, and J. S. Moodera, Nature materials **14**, 473 (2015).

[3] F. Arnold, C. Shekhar, S.-C. Wu, Y. Sun, R. D. Dos Reis, N. Kumar, M. Naumann, M. O. Ajeesh, M. Schmidt, A. G. Grushin, *et al.*, Nature communications **7**, 1 (2016).

[4] X. Yuan, C. Zhang, Y. Zhang, Z. Yan, T. Lyu, M. Zhang, Z. Li, C. Song, M. Zhao, P. Leng, *et al.*, Nature communications **11**, 1 (2020).

[5] C. Shekhar, F. Arnold, S.-C. Wu, Y. Sun, M. Schmidt, N. Kumar, A. G. Grushin, J. H. Bardarson, R. D. dos Reis, M. Naumann, *et al.*, Preprint at http://arxiv.org/abs/1506.06577 **21** (2015).

[6] J. Gooth, A. C. Niemann, T. Meng, A. G. Grushin, K. Landsteiner, B. Gotsmann, F. Menges, M. Schmidt, C. Shekhar, V. Süß, *et al.*, Nature **547**, 324 (2017).

[7] J. Gooth, B. Bradlyn, S. Honnali, C. Schindler, N. Kumar, J. Noky, Y. Qi, C. Shekhar, Y. Sun, Z. Wang, *et al.*, Nature **575**, 315 (2019).

[8] B. Bradlyn, J. Cano, Z. Wang, M. Vergniory, C. Felser, R. J. Cava, and B. A. Bernevig, Science **353**, aaf5037 (2016).

[9] J. Smit, Phys. Rev. B **8**, 2349 (1973).

[10] L. Berger, Physical Review B **2**, 4559 (1970).

[11] T. Jungwirth, Q. Niu, and A. MacDonald, Physical review letters **88**, 207208 (2002).

[12] M. Onoda and N. Nagaosa, Physical review letters **90**, 206601 (2003).

[13] Y. Yao, L. Kleinman, A. MacDonald, J. Sinova, T. Jungwirth, D.-s. Wang, E. Wang, and Q. Niu, Physical review letters **92**, 037204 (2004).

[14] J. Noky, Y. Zhang, J. Gooth, C. Felser, and Y. Sun, npj Computational Materials **6**, 1 (2020).

[15] E. H. Hall *et al.*, American Journal of Mathematics **2**, 287 (1879).

[16] S. N. Guin, K. Manna, J. Noky, S. J. Watzman, C. Fu, N. Kumar, W. Schnelle, C. Shekhar, Y. Sun, J. Gooth, *et al.*, NPG Asia Materials **11**, 1 (2019).

[17] A. Sakai, Y. P. Mizuta, A. A. Nugroho, R. Sihombing, T. Koretsune, M.-T. Suzuki, N. Takemori, R. Ishii, D. Nishio-Hamane, R. Arita, *et al.*, Nature Physics **14**, 1119 (2018).

[18] Z. Wang, Y. Sun, X.-Q. Chen, C. Franchini, G. Xu, H. Weng, X. Dai, and Z. Fang, Physical Review B **85**, 195320 (2012).

[19] K. Deng, G. Wan, P. Deng, K. Zhang, S. Ding, E. Wang, M. Yan, H. Huang, H. Zhang, Z. Xu, *et al.*, Nature Physics **12**, 1105 (2016).

[20] S.-M. Huang, S.-Y. Xu, I. Belopolski, C.-C. Lee, G. Chang, T.-R. Chang, B. Wang, N. Alidoust, G. Bian, M. Neupane, *et al.*, Proceedings of the National Academy of Sciences **113**, 1180 (2016).

[21] Z. Liu, B. Zhou, Y. Zhang, Z. Wang, H. Weng, D. Prabhakaran, S.-K. Mo, Z. Shen, Z. Fang, X. Dai, *et al.*, Science **343**, 864 (2014).

[22] H. Weng, C. Fang, Z. Fang, B. A. Bernevig, and X. Dai, Physical Review X **5**, 011029 (2015).

[23] S.-Y. Xu, I. Belopolski, N. Alidoust, M. Neupane, G. Bian, C. Zhang, R. Sankar, G. Chang, Z. Yuan, C.-C. Lee, *et al.*, Science **349**, 613 (2015).

[24] B. Lv, H. Weng, B. Fu, X. P. Wang, H. Miao, J. Ma, P. Richard, X. Huang, L. Zhao, G. Chen, *et al.*, Physical Review X **5**, 031013 (2015).

[25] N. Kumar, M. Yao, J. Nayak, M. G. Vergniory, J. Bannies, Z. Wang, N. B. M. Schröter, V. N. Strocov, L. Müchler, W. Shi, E. D. L. Rienks, J. L. Mañes, C. Shekhar, S. S. P. Parkin, J. Fink, G. H. Fecher, Y. Sun, B. A. Bernevig, and C. Felser, Advanced Materials **32**, 1906046 (2020), https://onlinelibrary.wiley.com/doi/pdf/10.1002/adma.201906046.

[26] N. B. M. Schröter, I. Robredo, S. Klemenz, R. J. Kirby, J. A. Krieger, D. Pei, T. Yu, S. Stolz, T. Schmitt, P. Dudin, T. K. Kim, C. Cacho, A. Schnyder, A. Bergara, V. N. Strocov, F. de Juan, M. G. Vergniory, and L. M. Schoop, Science Advances **6**, eabd5000 (2020), https://www.science.org/doi/pdf/10.1126/sciadv.abd5000.

[27] H. Schmid, Bulletin of Materials Science **17**, 1411 (1994).

[28] N. A. Hill, The journal of physical chemistry B **104**, 6694 (2000).

[29] J. B. Goodenough, Physical Review **100**, 564 (1955).

[30] J. Philipp, P. Majewski, L. Alff, A. Erb, R. Gross, T. Graf, M. Brandt, J. Simon, T. Walther, W. Mader, *et al.*, Physical Review B **68**, 144431 (2003).

[31] H. Tanaka and M. Misono, Current Opinion in Solid State and Materials Science **5**, 381 (2001).

[32] J. Zhu, H. Li, L. Zhong, P. Xiao, X. Xu, X. Yang, Z. Zhao, and J. Li, Acs Catalysis **4**, 2917 (2014).

[33] J. B. Goodenough and J.-S. Zhou, Journal of Materials Chemistry **17**, 2394 (2007).

[34] T. Chakraborty and S. Elizabeth, Journal of Magnetism and Magnetic Materials **462**, 78 (2018).

[35] M. Eibschütz, S. Shtrikman, and D. Treves, Physical review **156**, 562 (1967).

[36] T. Yamaguchi, Journal of Physics and Chemistry of Solids **35**, 479 (1974).

[37] M. Lorenz, M. R. Rao, T. Venkatesan, E. Fortunato, P. Barquinha, R. Branquinho, D. Salgueiro, R. Martins, E. Carlos, A. Liu, *et al.*, Journal of Physics D: Applied Physics **49**, 433001 (2016).

[38] M. Freedman, A. Kitaev, M. Larsen, and Z. Wang, Bulletin of the American Mathematical Society **40**, 31 (2003).

[39] T. Schuster, T. Iadecola, C. Chamon, R. Jackiw, and S.-Y. Pi, Physical Review B **94**, 115110 (2016).

[40] S. Murakami, N. Nagaosa, and S.-C. Zhang, Science **301**, 1348 (2003).

[41] S. Frolov, M. Manfra, and J. Sau, Nature Physics **16**, 718 (2020).

[42] Y. Chen, Y.-M. Lu, and H.-Y. Kee, Nature communications **6**, 1 (2015).

[43] J.-M. Carter, V. V. Shankar, M. A. Zeb, and H.-Y. Kee, Physical Review B **85**, 115105 (2012).

[44] S.-T. Pi, H. Wang, J. Kim, R. Wu, Y.-K. Wang, and C.-K. Lu, The journal of physical chemistry letters **8**, 332 (2017).

[45] N. C. Frey, M. K. Horton, J. M. Munro, S. M. Griffin, K. A. Persson, and V. B. Shenoy, Science advances **6**, eabd1076 (2020).

[46] R. Yu, H. Weng, Z. Fang, X. Dai, and X. Hu, Physical review letters **115**, 036807 (2015).

[47]S. Itoh, Y. Endoh, T. Yokoo, S. Ibuka, J.-G. Park, Y. Kaneko, K. S. Takahashi, Y. Tokura, and N. Nagaosa, Nature Communications **7**, 1 (2016).

[48]K.-I. Kobayashi, T. Kimura, Y. Tomioka, H. Sawada, K. Terakura, and Y. Tokura, Physical Review B **59**, 11159 (1999).

[49]Y. Tomioka, T. Okuda, Y. Okimoto, R. Kumai, K.-I. Kobayashi, and Y. Tokura, Physical Review B **61**, 422 (2000). [50]K.-I. Kobayashi, T. Kimura, H. Sawada, K. Terakura, and Y. Tokura, Nature **395**, 677 (1998).

[51]G. Kresse and J. Furthmüller, Phys. Rev. B **54**, 11169 (1996).

[52]G. Kresse and D. Joubert, Phys. Rev. B **59**, 1758 (1999).

[53]P. E. Blöchl, Phys. Rev. B **50**, 17953 (1994). [54]J. P. Perdew, K. Burke, and M. Ernzerhof, Phys. Rev. Lett. **77**, 3865 (1996).

[55]V. I. Anisimov, F. Aryasetiawan, and A. I. Lichtenstein, Journal of Physics: Condensed Matter **9**, 767 (1997).

[56]F. Aryasetiawan, M. Imada, A. Georges, G. Kotliar, S. Biermann, and A. I. Lichtenstein, Phys. Rev. B **70**, 195104 (2004). [57]A. Neroni, E. Şaşıoğlu, H. Hadipour, C. Friedrich, S. Blügel, I. Mertig, and M. Ležaić, Phys. Rev. B **100**, 115113 (2019). [58]X. Wang, J. R. Yates, I. Souza, and D. Vanderbilt, Phys. Rev. B **74**, 195118 (2006).

[59]N. Marzari, A. A. Mostofi, J. R. Yates, I. Souza, and D. Vanderbilt, Rev. Mod. Phys. **84**, 1419 (2012).

[60]I. Souza, N. Marzari, and D. Vanderbilt, Phys. Rev. B **65**, 035109 (2001).

[61]A. A. Mostofi, J. R. Yates, G. Pizzi, Y.-S. Lee, I. Souza, D. Vanderbilt, and N. Marzari, Computer Physics Communications **185**, 2309 (2014).

[62]Y. Yao, L. Kleinman, A. MacDonald, J. Sinova, T. Jungwirth, D.-s. Wang, E. Wang, and Q. Niu, Physical review letters **92**, 037204 (2004).

[63]A. Jain, S. P. Ong, G. Hautier, W. Chen, W. D. Richards, S. Dacek, S. Cholia, D. Gunter, D. Skinner, G. Ceder, and K. A. Persson, APL Materials **1**, 011002 (2013).

[64]A. Arulraj, K. Ramesha, J. Gopalakrishnan, and C. Rao, Journal of Solid State Chemistry **155**, 233 (2000).

[65]M. García-Hernández, J. Martínez, M. Martínez-Lope, M. Casais, and J. Alonso, Physical Review Letters **86**, 2443 (2001).

[66]C. Ang, Z. Yu, and L. Cross, Physical Review B **62**, 228 (2000).

[67]O. De Lima, J. Coaquira, R. De Almeida, L. De Carvalho, and S. Malik, Journal of Applied Physics **105**, 013907 (2009). [68]W. Tong, B. Zhang, S. Tan, and Y. Zhang, Physical Review B **70**, 014422 (2004).

[69]Z. Yu, C. Ang, P. Vilarinho, P. Mantas, and J. Baptista, Journal of applied physics **83**, 4874 (1998).

[70]O. Bidault, P. Goux, M. Kchikech, M. Belkaoumi, and M. Maglione, Physical Review B **49**, 7868 (1994). [71]J. Noky, Q. Xu, C. Felser, and Y. Sun, Physical Review B **99**, 165117 (2019).

[72]I. Robredo, N. Schröter, A. Reyes-Serrato, A. Bergara, F. de Juan, L. M. Schoop, and M. G. Vergniory, Journal of Physics D: Applied Physics **55**, 304004 (2022).

[73]L. E. P. S. e. a. Sohn, B., Nat. Mater. **20** (2021).